\documentclass[conference]{IEEEtran}

\usepackage[dvips, final]{graphicx}
\usepackage{array}
\usepackage{amsmath}
\usepackage{amssymb}
\usepackage{amsfonts}
\usepackage{graphicx}
\usepackage{epsfig}
\usepackage{psfrag}

\def\b0{{\mathbf 0}}
\def\bA{{\mathbf A}}
\def\bB{{\mathbf B}}

\def\bh{{\mathbf h}}

\def\bq{{\mathbf q}}
\def\bs{{\mathbf s}}

\def\bx{{\mathbf x}}

\def\by{{\mathbf y}}
\def\bz{{\mathbf z}}

\def\bH{{\mathbf H}}
\def\bU{{\mathbf U}}
\def\bI{{\mathbf I}}

\def\bQ{{\mathbf Q}}
\def\bV{{\mathbf V}}

\def\bG{{\mathbf G}}

\def\bX{{\mathbf X}}

\title{On Two-way Communications for \\ Cooperative Multiple Source Pairs \\ Through a Multi-antenna Relay}

\author{\authorblockN{Chin Choy CHAI\authorrefmark{1},
Chau YUEN\authorrefmark{2}
\vspace{3pt}}
\authorblockA{\authorrefmark{1}Institute for Infocomm Research \\
1 Fusionopolis Way, \#21-01 Connexis, Singapore 138632 \\
\small E-mail: chaicc@i2r.a-star.edu.sg}

\authorblockA{\authorrefmark{2}Singapore University of Technology and Design \\
287 Ghim Moh Road \#04-00, Singapore 279623 \\}\vspace{1pt}
\small E-mail: yuenchau@sutd.edu.sg}

\begin{document}
\maketitle

\begin{abstract}

We study amplified-and-forward (AF)-based two-way relaying (TWR) with
multiple source pairs, which are
exchanging information through the relay.  Each source has
single antenna and the relay has multi-antenna.  The optimal beamforming matrix structure that achieves
maximum signal-to-interference-plus-noise ratio (SINR) for TWR with multiple source pairs is derived.  We then present two new non-zero-forcing based beamforming schemes for TWR, which take into consideration the tradeoff between preserving the desired signals
and suppressing inter-pair interference between different source pairs.  Joint grouping and beamforming scheme is proposed to achieve a better signal-to-interference-plus-noise ratio (SINR) when the total number of source pairs is large and the signal-to-noise ratio (SNR) at the relay is low.
\end{abstract}


\setlength{\baselineskip}{1.3\baselineskip}
\newtheorem{lemma}{\underline{Lemma}}[section]
\newtheorem{remark}{\underline{Remark}}[section]
\newtheorem{corollary}{\underline{Corollary}}[section]
\newtheorem{problem}{\underline{Problem}}[section]
\newtheorem{algorithm}{\underline{Algorithm}}[section]
\newcommand{\mv}[1]{\mbox{\boldmath{$ #1 $}}}

\section{Introduction}\label{sec:intro}
By applying the physical-layer network coding \cite{PHY_network coding} or
analogue network coding (ANC) \cite{ANC} in  two-way relaying (TWR), only two time slots
are required for one complete information exchange using TWR.  In the first
time slot, both source nodes transmit simultaneously to the relay.
In the second time slot, the relay broadcasts the common message
which is obtained by combining the received messages.  Since both source nodes know their own transmitted signals, each of their
self-interference can be completely canceled prior to decoding.

The TWR has been studied in \cite{Hammerstrom et al 2007} to
\cite{Rui Zhang Ying-Chang Liang Chin Choy Chai  and Shuguang Cui 2009}  for
the case of single source pair. The beamformer
design for AF-based multiple-input-multiple-output (MIMO) TWR is studied in
\cite{Unger2007}\cite{Timo Unger and Anja Klein 2008}, in which the
receive and transmit beamforming are derived separately and then
combined to form the relay beamformer.  Furthermore, some strategies to enhance the performance of TWR can be found in \cite{Eslamifar2010} to \cite{Song Jul2010} and references therein.

The non-ANC-based TWR for multiple source pairs is studied in \cite{Abe Shi Asai and Yoshino 2006} to \cite{Esli and Wittneben 2008_2}.  Unlike the single source pair case, in multiple source pairs scenario, additional
inter-pair interference exists between different
source pairs, which degrades the TWR performance.  In \cite{Abe Shi Asai and Yoshino 2006}, a relay network with
multiple source pairs and multiple relay nodes is studied, where all
sources and relay stations have multiple antennas. The multiuser TWR is proposed and studied in \cite{Rankov07}, where multiple source pairs are communicating via multiple relays. In \cite{Esli and Wittneben 2008} \cite{Esli and Wittneben 2008_2}, the MIMO TWR where multiple wireless node pairs are communicating via a single decode-and-forward (DF) relay is studied.

Due to the presence of inter-pair interference, previous beamforming
solutions for TWR with single source pair is no longer useful and
new solutions are required for the case of multiple source pairs.
In almost all the above works \cite{Abe Shi Asai and Yoshino 2006} to \cite{Esli and Wittneben 2008_2}, the
inter-pair interference are canceled using zero-forcing (ZF)-based approach.  

In this paper, two new non-ZF-based beamforming schemes or beamformers are proposed for ANC-based TWR. Instead of completely canceling the inter-pair interference for all source pairs by ZF-based methods as was done in previous works, we propose joint grouping and beamforming scheme that divides a given large number of source pairs into smaller subgroups, and then apply the proposed beamformers to each subgroup.  To the best of our knowledge, this approach has not been studied in previous works.  Simulation results are presented to compare the performance of the proposed schemes.

\section{System Model}\label{sec:ANC-MAR}

We study wireless TWR with a multi-antenna relay and a total number of $K_T$ single-antenna sources, where $K_T$ is an even number.  Due to the fact that the inter-pair interference contains the desired signals of all the other sources, these desired signals are also suppressed by any suboptimal beamformer which causes significant loss in the SINR, especially when $K_T$ is large.  We propose to overcome that this shortcoming by first dividing a large number of $K_T$ source pairs into $N$ subgroups, each with a smaller number of $K$ source pairs, where $K$ is an even number.  Then, by using time division approach, the relay performs non-ZF-based beamforming on each subgroup of users at one time, and take turn to serve all source pairs, to achieve a better throughput performance.  Next, we consider a given subgroup of $K$ sources, and derive TWR beamformers for these $K$ sources.

Without loss of
generality, we assume the $k$-th source node $S_{k}$,
$k=\hdots,K$, is to exchange information with another source $S_{\tilde
k}$, where $\tilde k=1, \hdots,K$, $k\ne \tilde k$. Each $k$-th
source have single antenna, whereas the relay station R
is equipped with $M (M \ge K-1)$ antennas.  Let the ${M\times1}$ vectors $\bh_{k}$ and $\bh_{\tilde k}$ denote, respectively, the channel response
matrices from $S_{k}$ to R, and that from $S_{\tilde k}$ to R.  We
assume that the elements of $\bh_{k}$ and $\bh_{\tilde k}$ follow
the distribution of circularly symmetric complex Gaussian with zero
mean and unity variance, which is denoted as
$\mathcal{CN}(0,1)$.  Let $s_{k}(n)$ and $s_{\tilde k}(n)$ denote respectively, the
transmitted symbols from $S_{k}$ to $S_{\tilde k}$, and from
$S_{\tilde k}$ to $S_{k}$.  We assume the optimal Gaussian codebook
is used at each $S_{k}$, and therefore $s_{k}(n)$'s are independent
random variable each is distributed as $\mathcal{CN}(0,1)$.

Assume TDD is used and two time slots are needed for information
exchange using analogue network coding (ANC)\cite{PHY_network
coding}, \cite{ANC}. In the first time slot, all active source nodes
transmit their signals simultaneously, the received baseband signal
vector $\by_R(n)$ at R is given by,
\begin{equation}
\begin{split}
\by_R(n) =& \bh_{1} {\sqrt p_1} s_1(n) + \hdots+ \bh_{K-1} {\sqrt p_{K-1}} s_{K-1}(n) \\
&+ \bh_{K} {\sqrt p_K} s_{K}(n)+ \bz_R(n), \label{eq:ANC-MAR-1}
\end{split}
\end{equation}
where $n$ is symbol index, $p_k$ is the transmit power at $S_k$,
$\bz_R(n)$ is the received additive
noise vector, and without loss of generality, it is assume that
$\bz_R(n)$ follows the distribution of
$\mathcal{CN}(0,\bI\sigma^{2}), \forall n$, where $\bI$ is an identity matrix.  Throughout this paper, we assume that the $p_k$'s are given or fixed.

At the relay, we consider AF relaying using linear beamforming which is
represented by a $M\times M$ matrix $\bA$. The
transmit signal $\bx_R(n)$ at R can be expressed in terms of its
inputs $\by_R(n)$ as $\bx_R(n) = \bA \by_R(n)$.  We assume channel reciprocity for uplink
and downlink transmission through the relay. In the second time slot, when
$\bx_R(n)$  is transmitted from R, the channels from R to $S_k$ become
$\bh_{k}^{T}$, $k=1, \hdots, K$.
The total transmit power at R, denoted as $p_R$, can
be shown as,
\begin{equation}
p_R=\sum_{k=1}^{K} \| \bA \bh_k \|^{2} p_{k}  + {\rm Tr} (\bA
\bA^{\dag}) \sigma^2, \label{eq:Relay power}
\end{equation}
where ${\rm Tr}(\bX)$ denotes the trace of $\bX$.  
%

The ANC is adopted as follows. We assume
that using training and estimation, $\bh_{k}^{T} \bA \bh_{k}$ and
$\bh_{k}^{T} \bA \bh_{\tilde k}$ are perfectly known at $S_k$,
$k=1,\hdots,K$ prior to signal transmission. Each of the $S_k$ can
first cancel its self-interference and then coherently demodulate
for $s_{\tilde k}$. This yields
\begin{equation}
\begin{split}
{\tilde y}_k (n) = & \bh_{k}^{T} \bA \bh_{\tilde k} {\sqrt p_{\tilde
k}} s_{\tilde k}(n)  + \sum_{j\ne k, j\ne {\tilde
k}}\bh_{k}^{T}\bA \bh_{j} {\sqrt p_j} s_j(n)\nonumber \\
& +{\tilde z}_{k}(n), k=1,\hdots, K, \label{eq:ANC-MAR-recS1}
\end{split}
\end{equation}
for $k=1,\hdots,K$.  We assume that the received noise $z_{k}(n)$ is distributed as
$\mathcal{CN}(0,\sigma^{2})$, and $\bz_R(n)$ are independent of
$z_k(n)$. ${\tilde z}_{k}(n)=\bh_{k}^{T} \bA \bz_R(n) + z_{k}(n)$, and
${\tilde z}_{k}(n)$ is distributed as $\mathcal{CN}(0, (\|\bh_k^{T}
\bA\|^2+1)\sigma^{2}) $.  At each $S_k$, coherent signal detection can then be used to recover $\bs_{\tilde k}(n)$
from ${\tilde y}_k (n)$.
The signal-to-interference-plus-noise ratio (SINR) for the $k$-th
destination node.\, $k=1,\hdots,K$, can be expressed as
\begin{equation}
\gamma_{k} = \frac{|\bh_{k}^{T} \bA \bh_{\tilde k}|^2 p_{\tilde k}
}{\sum_{j\ne k, j\ne {\tilde k}} |\bh_{k}^{T} \bA \bh_{j}|^2 p_{j}(n)+
(\|\bh_k^{T} \bA\|^2 + 1)\sigma^{2}}.
\label{eq:SINRk} \\
\end{equation}\\

\section{Proposed Schemes}
\subsection{Optimal beamformer}
We define the uplink (UL) channel gain matrix $\bH_{UL} = [\bh_{1},
\bh_{2},\hdots, \bh_{K}]$ and denote the singular value
decomposition (SVD) of $\bH_{UL}$ as $\bH_{UL} = \bU_{} {\mathbf \Sigma}_{}
\bV_{}^{H}$, where the $M\times K$ matrix $\bU_{}$, and the $K\times K$ matrix $\bV_{}$, are with orthogonal column vectors, and
${\mathbf \Sigma}_{}$ is singular value matrix with dimension $K
\times K$ where $[{\mathbf \Sigma}_{}]_{\ell,\ell} = \sigma_{\ell}$
for $\ell =
1, \hdots,K$. We have the following new result.


\emph{Proposition:} The optimal beamforming matrix to achieve
maximum SINR in (\ref{eq:SINRk}) has the following structure:
\begin{eqnarray}
\bA^{\rm opt} = \bU_{}^{*} \bB^{\rm opt} \bU_{}^{H},
\label{eq:optimalBAF}
\end{eqnarray}
where $\bB^{\rm opt}$ is a $K\times K$ matrix.

\noindent \emph{Proof:} The above new result can be proven by extending the proof of \cite{Liang
and Zhang 2008}\cite{Rui Zhang Ying-Chang Liang Chin Choy Chai  and Shuguang Cui 2009}, which considers the TWR for the case of single source pair who are exchanging information.  For the case of multiple source
pairs, each source is also subject to interference from the other
source pairs. This so-called inter-pair interference term also
depends only on $\bB$ as $\sum_{j\ne
k, j\ne {\tilde k}}|\bh_{k}^{T} \bA \bh_{j}|^2=\sum_{j\ne k, j\ne
{\tilde k}} |\bh_{k}^{T} \bU_{}^{*}\bB\bU_{}^{H} \bh_{j}|^2$, which
also spans the total signal subspace
of $\bB$.   Therefore, we obtain $\bB^{\rm opt}$.  The beamforming matrix $\bB$ can be solved as shown next.  \\

Let ${\tilde \bh}_{k}=\bU_{}^{H}\bh_{k}$, $k=1,\hdots,K$,  represent the effective channel from $S_k$ to
R, with ${\bf A}$ given in (\ref{eq:optimalBAF}). Similarly, let
${\tilde \bh}_{k}^{T}$ represent the effective channel from R to
$S_k$.  The SINR formula in (\ref{eq:SINRk}) can be written in terms of ${\bf B}$.  Throughout this paper, the optimization problems are formulated in
terms of the beamforming matrix $\bB$ and the effective
channels ${\tilde \bh}_{k}$.

The minimum power (MP) beamformer is derived
by minimizing the total relay transmit power with respect to the
relay beamforming matrix ${\bf B}$ (or equivalently, ${\bf A}$), subject to SINR constraints
$\gamma_k \ge \gamma_k^{\ast}$, $k=1, \hdots, K$,
\begin{equation}\label{eq:optimization1}
\begin{split}
\min_{\bf B} & ~~\sum_{k=1}^{K} \|{\bB}{{\tilde \bh}_k}\|^{2} p_{k}+{\rm Tr} (\bB\bB^{H})\sigma^2,\\
\mbox{subject to} & ~~\gamma_{k} \ge \gamma_k^{\ast}, k=1,\hdots,K.
\end{split}
\end{equation}

For given transmit powers $p_{k}$, $k=1, \hdots, K$, we can use the same approach as in \cite{Rui Zhang Ying-Chang Liang Chin Choy Chai  and Shuguang Cui 2009}, to develop an efficient algorithm based on the second-order cone programming (SOCP) \cite{Boydbook} to solve the problem in \eqref{eq:optimization1}.

\subsection{Suboptimal beamformer}
We derive the suboptimal minimum interference (MI) beamformer that does not
require computations via optimization technique.  The tradeoff
between the desired signals and interference is taken into account
by minimizing the sum of inter-pair interference plus AF noise at
the output of the beamformer, subject to the constraints that the
desired signal gain for each $k$-th receiver is equal to a constant
$\beta_k$. The additive Gaussian noise ${\hat z}_{k}(n)$ at each
$k$-th source, which is not affected by the beamforming matrix ${\bf B}$,
and is neglected in the optimization. Again in this case, the
joint design of both receive and transmit beamforming is considered.
The interference minimization problem is
formulated as,
\begin{equation}
\begin{split}
  \min_{{\bf B}} &~~ \sum_{k=1}^K I_k({\bf B}),   \\
\mbox{subject to}&~~|{\tilde \bh}_k^T{\bf B}{\tilde \bh}_{\tilde
k}|^2 p_{\tilde k}=\beta_k, k=1, \hdots ,K,
\label{eq:MinimizeInterference}
\end{split}
\end{equation}

\noindent where
\begin{equation}
\begin{split}
I_k({\bf B})=& \sum_{j\ne k, j\ne {\tilde k}} |{\bh}_k^T{\bf
B}{\tilde \bh}_j|^2 p_j + \|{\tilde \bh}_k^{T} \bB \|^2\sigma^2, k=1, \hdots
,K.
\label{eq:Interference}
\end{split}
\end{equation}

The above constraints in (\ref{eq:MinimizeInterference}) are
introduced to preserve the desired signal components at each source,
so as to minimize the inter-pair interference plus
AF noise component that point into any undesired direction. In
(\ref{eq:Interference}), $I_k({\bf B})$ represents the sum of inter-pair interference (from
other source pairs) plus AF noise that is imposed on $S_k$.  To solve for ${\bf B}$, the UL and DL
channel response of each $k$-th source is assumed to be known at the
relay.

We assume that the $p_k$'s are given, and write $|{\tilde \bh}_{k}^{T} \bB
{\tilde \bh}_{\tilde k} |^2p_{\tilde k}=| {\bf f}_k^T \bf b|^2 $,
$\sum_{j\ne k, j\ne {\tilde k}}|{\tilde \bh}_{k}^{T} \bB {\tilde
\bh}_{j}|^2 p_{j}=\sum_{j\ne k, j\ne {\tilde k}} | {\bf d}_{k,j}^T
{\bf b}|^2$, and $\|{\tilde \bh}_k^{T} \bB \|^2=\| {\bG_k} {\bf
b}\|^2$, where $k \ne {\tilde k}$, and ${\bf b}$ denotes a $K^2 \times 1$ equivalent
beamforming weights vector which is generated by the rule of
(\ref{eqn:Vector}) as,
\begin{equation}
    \label{eqn:Vector}
       \mathcal{V}(\bQ)= \left[ \begin{array}{c}
        \bq_1  \\
         \vdots\\
          \bq_K
        \end{array} \right].
    \end{equation}
The problem (\ref{eq:MinimizeInterference}) can be written in terms of
${\bf b}$ as,
\begin{equation}
\begin{split}
  \min_{{\bf b}} &~~\sum_{k=1}^K ( \sum_{j\ne k, j\ne {\tilde k}} |{\bf d}_{k,j}^T {\bf b}|^2 +\| {\bf G}_k {\bf
b}\|^2 \sigma^2)\\
\mbox{subject to}&~~| {\bf f}_k^T {\bf b}|=\beta_k,
k=1,\hdots,K.\label{eq:MinimizeInterference2}
\end{split}
\end{equation}

We further define ${\bf R}_{k}= \sum_{j\ne k, j\ne {\tilde k}}{\bf
d}_{k,j}^{\ast} {\bf d}_{k,j}^T$, ${\bf N}_{k}={\bf G}_k^{H}{\bf
G}_k$, where both ${\bf R}_{k}$, and ${\bf N}_{k}$ are $K^2 \times
K^2$ matrices.  By writing $\| {\bf G}_k {\bf b}\|^2={\bf b}^{H}{\bf
N}_{k}{\bf b}$, $\sum_{j\ne k, j\ne {\tilde k}}|{\bf d}_{k,j}^T
{\bf b}|^2={\bf b}^{H}{\bf R}_{k}{\bf b}$, and ${\bf \Phi}=\sum_{k=1}^K ({\bf R}_{k}+{\bf
N}_{k} \sigma^2 )$, the problem
(\ref{eq:MinimizeInterference2}) can be written as,
\begin{equation}
\begin{split}
\min_{{\bf b}} &~~{\bf b}^{H}{\bf \Phi} {\bf b}, \\
\mbox{subject to}&~~{\bf C}^H {\bf b} = {\bf g}.\label{eq:LCMV}
\end{split}
\end{equation}
where ${\bf C}=[{\bf f}_1^{\ast} ~{\bf
f}_2^{\ast}~ {\bf f}_3^{\ast}~ {\bf f}_K^{\ast} ]$ is the $K^2 \times K$ constraint
matrix, and ${\bf f}_k$ has been defined in the paragraph after
Eq.(\ref{eqn:Vector}).  Each of the
column of ${\bf C}$ imposes a constraint on the equivalent beamforming weight vector ${\bf b}$.  The $K \times 1$ response vector ${\bf
g}=[\beta_1~\beta_2~ \cdots \beta_K]^T$ contains the K scalar
constraints values $\beta_k, k=1,\hdots,K$, which are chosen to
satisfy the SINR requirements of each source.

    The beamforming vector solution, denoted as ${\bf b}_{\rm MI}$, that
satisfies (\ref{eq:LCMV}) is solved as \cite{Frost 1972},
\begin{equation}
\begin{split}
 {\bf b}_{\rm MI}= {\bf \Phi}^{-1}{\bf C}({\bf C}^{H}{\bf \Phi}^{-1} {\bf C}) ^{-1}{\bf g}.\label{eq:LCMVsolution}
\end{split}
\end{equation}
The MI relay beamformer, denoted as ${\bf B}_{\rm MI}$, can then be
obtained from ${\bf b}_{\rm MI}$ as ${\bf B}_{\rm MI}
=\mathcal{V}^{-1}(\alpha {\bf b}_{\rm MI})$, where $\mathcal{V}^{-1}$ denotes the inverse operation of $\mathcal{V}$ defined in  \eqref{eqn:Vector}.  It can be shown that the
solution ${\bf B}_{\rm MI}$ maximizes the SINR of
each source, where the constant $\alpha$
is used to control the total relay power $p_R$ in (\ref{eq:Relay power}).
The iterative steps to search for $\alpha$ are presented next.

\subsection{Joint grouping and beamforming scheme}
We propose the following joint grouping and beamforming scheme that divides a given large number of source pairs into smaller subgroups, and then apply the above beamformers to each subgroup.  For simplicity, here we consider that the grouping is done arbitrarily.  By doing so we can reduce the feedback of channel state information and beamforming calculation during each relaying.  

\begin{itemize}
\item Divide the total number of $K_T$ source pairs into $N$ subgroups, $N=1,2,\hdots$, and let $K=K_T/N$ denote the number of source in each subgroup.
\item For each $n$-th subgroup, $n=1, \hdots, N$, compute the MI beamforming matrix as follows:
\item {\bf Given} $\alpha\in [0, \alpha_{\rm max}]$, ${\bf p}=[p_1~p_2~ \ldots ~ p_K]^{T}$.
\item {\bf Initialize} $\alpha_{\rm lower}=0$, $\alpha_{\rm upper}=\alpha_{\rm max}$.

\item Compute the SVD of the UL channel matrix $\bH_{UL} = \bU_{} {\mathbf
\Sigma}_{} \bV_{}^{H}$.
\item Estimate the correlation matrix ${\bf \Phi}$.
\item Compute the constraint matrix ${\bf C}$ and the response vector ${\bf g}$.

\item Compute the beamforming weights using ${\bf b}_{\rm MI}=$ ${\bf \Phi}^{-1}{\bf C}({\bf C}^{H}{\bf \Phi}^{-1} {\bf C}) ^{-1}{\bf
g}$.

\item Obtain the MI relay beamforming matrix by arranging the beamforming weights of ${\bf b}_{\rm MI}$ into the matrix form, ${\bf B}_{\rm MI}
=\mathcal{V}^{-1}(\alpha{\bf b}_{\rm MI})$.

\item {\bf Repeat}
\begin{itemize}
\item[1.] Set $\alpha \leftarrow\frac{1}{2}(\alpha_{\rm lower}+\alpha_{\rm upper})$.
\item[2.] Update $\alpha$ by the bisection method \cite{Boydbook}: If
$p_R$ is less than a given power constraint, set $\alpha_{\rm lower}\leftarrow \alpha$;
otherwise, $\alpha_{\rm upper} \leftarrow \alpha$.
\end{itemize}
\item {\bf Until} $\alpha_{\rm upper}-\alpha_{\rm lower}\leq\delta_{\alpha}$, where the small positive constant
$\delta_{\alpha}$ is chosen to ensure sufficient accuracy.
\item Compute the MI relay beamforming matrix as $\bA_{\rm MI} = \bU_{}^{*} \bB_{\rm MI} \bU_{}^{H}$ for each $n$-th subgroup, $n=1, \hdots, N$.
\item Select the number of subgroup $N$ and its corresponding $\bA_{\rm MI}$ which results in the largest achievable sum-rate.\\
\end{itemize}

\section{Simulation results and discussions} \label{sec:numerical results}
In this section, the sum rate performance of the proposed TWR
beamformers are presented.  For simplicity, the channel correlations between the $j$-th and the $k$-th
sources are set to be equal, that is, $|{\bf h}_j^H{\bf h}_k|^2=\rho_{}$, $\forall j \ne k$.  

\begin{figure}[!htb] \centerline{\includegraphics[width=1\columnwidth]{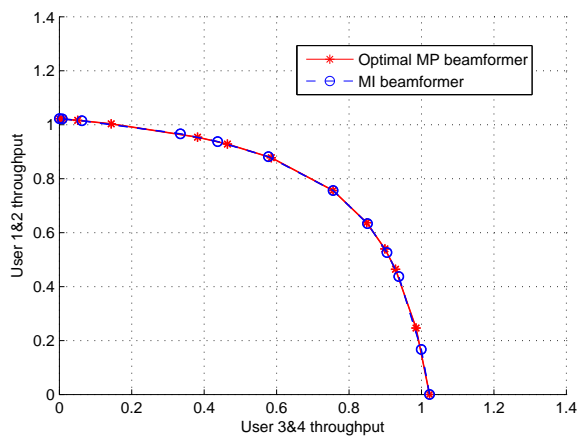}}
\caption{Comparison on the achievable rate region for the proposed MP and the MI relay beamformers with $M=8$, $K=4$, fixed $p_1=p_2=p_3=p_4=10$, $P_R=10$, and $\rho=0$.}\label{fig:rate comp no cor}
\end{figure}

In Fig. \ref{fig:rate comp no cor}, two pairs of single-antenna source ($K=4$) and a multi-antenna relay ($M=8$) are considered in
the simulations. The transmit power at each source $S_1$, $S_2$,
$S_3$ and $S_4$ are fixed as $p_1=p_2=p_3=p_4=10{\rm dB}$, and the
relay transmit power is set to be $p_R=10 {\rm dB}$.  Both sources within each pair are set to have
identical SINR requirement.  With no channel correlation between different sources, the proposed MI beamformer achieves the same achievable rate region as the optimal MP beamforming scheme.  However, when the channel correlations between different source pairs, and within each source pair, are set to be higher, the proposed MI beamformer achieves a smaller rate region (not shown here due to space constraint).  This is because when the channels of different source pairs are correlated, the desired signal components that point in the direction of the inter-pair interference will also be suppressed by the MI beamformer, which in
turn reduces the achievable rate for each source.

\begin{figure}[!htb] \centerline{\includegraphics[width=1\columnwidth]{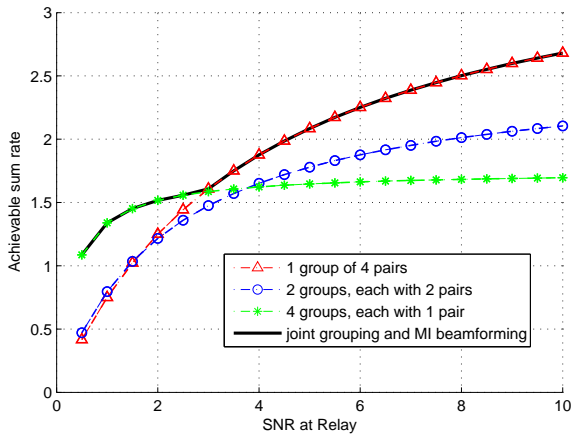}}
\caption{Comparison on the sum-rate versus SNR for the proposed MI relay beamformer with
$M=8$, $K=8$, $\rho=0$.}\label{fig:sum-rate no cor}
\end{figure}

In Fig. \ref{fig:sum-rate no cor}, we present the achievable sum-rate versus SNR at the relay, which is defined as $p_{R}/\sigma^2$, for the proposed MI beamformer with various number of subgroups.  Four pairs of single-antenna source ($K=8$) and a multi-antenna relay ($M=8$) are considered, and the channels between different sources are uncorrelated.  Time division is used to serve different subgroup.  We observe that either the smaller subgroups or large group transmission does not always perform the best in TWR.  With low SNR at the relay, the case of four groups (each with one source pair) performs the best, whereas for large SNR, the case of a single group (with four source pairs) performs the best.  This is because with either large number of source pairs or small relay transmit power, the MI beamformer tends to suppress the interference more for the case of single group (with more interfering users), which results in SINR loss.

 To overcome this shortcoming, joint grouping and beamforming scheme can be used to reduce SINR loss as follows.  For small SNR, we should apply the proposed beamformers to four different subgroups, and for large SNR, we should apply the proposed beamformers to a single group (with four source pairs).  The improved sum-rate performance by joint grouping and MI beamforming scheme is highlighted by the solid line in Fig. \ref{fig:sum-rate no cor}.  The optimal grouping and selection of the SNR thresholds correspond to different number of subgroups are interesting subjects for further study. \\ 

\section{Conclusions}\label{sec:conclusions}
New optimal and suboptimal beamformers for
ANC-based TWR with multiple source pairs are derived by taking into account the tradeoff between the desired signals and the inter-pair interference.  For low SNR, a better sum-rate performance can be achieved by first diving a large number of source pairs into smaller subgroups, and then apply beamforming to each subgroup using time division.\\  

\section*{Acknowledgment}
This research is partly supported by the Singapore University Technology and Design (grant no. SRG-EPD-2010-005).

\bibliographystyle{ieeetr}

\newpage
\linespread{1.45}

\end{document}